\documentclass[%
 aip,
rsi,%
 amsmath,amssymb,
 preprint,%
]{revtex4-1}

\usepackage{graphicx}
\usepackage{dcolumn}
\usepackage{bm}
\usepackage[utf8]{inputenc}
\usepackage{mathrsfs}
\usepackage{algorithm}
\usepackage{algpseudocode}
\usepackage{xcolor}
\usepackage[version=4]{mhchem}

\newcommand{\bsym}{\boldsymbol} 

\renewcommand{\b}{\bsym}

\newcommand{\ket}[1]{\vert #1 \rangle}

\newcommand{\braket}[1]{\langle #1 \rangle}

\newcommand{\Tbraket}[3]{\langle #1 \hspace{.10em} \vert \hspace{.10em} #2 \hspace{.10em} \vert \hspace{.10em} #3 \rangle}

\newcommand{\J}{\bsym{A}}

\newcommand{\norm}[1]{\| #1 \|}

\newcommand{\hf}{\mathrm{HF}}
\newcommand{\cc}{\mathrm{CC}}

\newcommand{\aux}{\mathscr{A}}
\newcommand{\target}{\mathscr{T}}
\renewcommand{\cc}{\Psi_0}
\newcommand{\order}[1]{\mathscr{O}(#1)}

\begin{document}

\title{Accelerated multimodel Newton-type algorithms for faster convergence of ground and excited state coupled cluster equations}

\author{Eirik F.~Kjønstad}
\affiliation{ 
Department of Chemistry, Norwegian University of Science and Technology, 7491 Trondheim, Norway
}%
\author{Sarai D.~Folkestad}
\affiliation{ 
Department of Chemistry, Norwegian University of Science and Technology, 7491 Trondheim, Norway
}%
\author{Henrik Koch}
\email{henrik.koch@sns.it}
\affiliation{%
Scuola Normale Superiore, Piazza dei Cavaleri 7, 56126 Pisa, Italy 
}%
\affiliation{ 
Department of Chemistry, Norwegian University of Science and Technology, 7491 Trondheim, Norway
}%

\date{\today}

\begin{abstract}
We introduce a 
multimodel  
approach
to solve 
coupled cluster equations, employing a quasi Newton algorithm for the ground state 
and an 
Olsen algorithm for the excited states.
In these algorithms, 
both of which can be viewed as Newton algorithms,
the Jacobian matrix of a lower level 
coupled cluster
model is used 
in 
Newton equations 
associated with
the target model.
Improvements in convergence
then implies 
savings for sufficiently large molecular 
systems, 
since
the  
computational cost of 
macroiterations  
scales more steeply
with system size
than the cost of 
microiterations. 
The 
multimodel
approach is 
suitable
when
there is a lower level Jacobian matrix 
that is 
much
more accurate 
than the zeroth order
approximation.
Applying
the approach to the CC3 equations, using the CCSD approximation of the Jacobian, 
we
show that the time spent to determine the ground and valence excited states can be significantly reduced. We also find 
improved convergence
for core excited states, indicating that similar savings 
will be obtained
with
an explicit implementation of the core-valence separated CCSD Jacobian transformation. 
\end{abstract}

\maketitle

\section{Introduction}
The molecular systems that can be treated with coupled cluster theory is severely limited by its steep computational scaling.\citep{Bartlett2007} 
When faced with this ``scaling wall,'' 
two complementary approaches can be pursued to extend the 
application range. 
One can either introduce approximations, as in linear scaling\citep{crawford2010reduced} and
multilevel\citep{Myhre2014, myhre2016multilevel, folkestad2019multilevel} coupled cluster methods, or optimize the evaluation and convergence of model equations.
The latter can give significant savings, given that for an iterative coupled cluster method with a computational cost of $\order{M^k}$, where $M$ is the number of molecular orbitals and $k \in \mathbb{N}$, 
each iteration involves evaluation of equations that scale as $\order{M^k}$. 
 
Currently,
the method of choice\citep{Helgaker2014} 
to converge
the ground state amplitude equations 
is
a quasi Newton algorithm\citep{dennis1977quasi} accelerated by direct inversion of the iterative subspace\citep{PULAY1980393,Pulay1982} (DIIS). 
In this Newton algorithm, 
the derivative of the residual---\emph{i.e.}, the coupled cluster Jacobian\citep{Koch1990}---
is approximated
by substituting the Hamiltonian with the zeroth order Fock matrix approximation. 
The DIIS acceleration technique was introduced by Pulay\citep{PULAY1980393,Pulay1982} to improve convergence of the Hartree-Fock equations, but it has subsequently been applied successfully in other areas of quantum chemistry, including geometry optimization\citep{CSASZAR198431,farkas2002methods} and coupled cluster theory.\citep{SCUSERIA1986236} 
In an iteration using DIIS, current and previous update estimates, \emph{e.g.}~obtained using a Newton algorithm, are combined to minimize an averaged error vector in the least-squares sense. Although standard DIIS is normally highly effective, the related {conjugate residual with optimal trial vectors (CROP)} method\citep{CROP2008,Ettenhuber2015} can be used to ensure that optimal vectors are kept in the DIIS subspace. 
{Another related acceleration method is the reduced linear equation (RLE) approach of Purvis and Bartlett.\citep{Purvis1981,TRUCKS1988548}}

The zeroth order 
coupled cluster Jacobian has a number of advantages. In many cases 
it approximates the exact Jacobian reasonably well,
leading to rapid convergence when combined with DIIS. 
In addition, it is a diagonal matrix consisting of orbital energy differences, implying an analytically solvable Newton equation: the amplitude update estimate is simply the residual vector weighted by inverse orbital energy differences. 
This makes it particularly easy to implement. Furthermore, 
the computational cost involved in weighting the residual vector is negligible, being of $\order{M^2}$ for perturbative doubles\citep{Christiansen1995CC2} (CC2) and $\order{M^4}$ for singles and doubles\citep{Purvis1982} (CCSD) as well as singles and doubles with perturbative triples\citep{Christiansen1995,Koch1997} (CC3).
This can be contrasted to 
the cost of evaluating the residual with these models, which scales as $\order{M^5}$,  $\order{M^6},$ and $\order{M^7}$, respectively. 

For the more expensive coupled cluster models 
in the hierarchy---especially those describing triples or higher excitations---it is natural to enquire whether improved convergence can be attained by making use 
of the lower levels in the hierarchy. 
In this paper we 
consider
this possibility by using 
quasi 
Newton algorithms in which the Jacobian matrix 
is approximated at a lower level of theory. We  refer to the lower level model as the \emph{auxilliary} model $\aux$ to distinguish it from the \emph{target} model $\target$. Within this 
multimodel Newton
framework, macroiterations have a cost of $\order{M^{k_\target}}$
and
microiterations 
a cost of $\order{M^{k_\aux}}$, where $k_\aux < k_\target$. 
The $\order{M^{k_\target}}$ terms begin to dominate
for large $M$. 
The computational cost can 
thus
reduce significantly if the number of macroiterations decrease.

Similar observations were recently made by Matthews and Stanton.\citep{Matthews2015} They found that the convergence of the amplitude equations in full triples\citep{Noga1987,Noga1988erratum} (CCSDT) and quadruples\citep{Kucharski1992} (CCSDTQ) 
could be 
improved
by performing ``subiterations'' in which only low-order amplitudes are updated. This also led to a lowering of the total computational cost.\citep{Matthews2015} Very recently, Matthews\citep{matthews2020accelerating} extended the approach in Ref.~\citenum{Matthews2015} to the left ground state equations.

Excited states are derived from a linear eigenvalue problem,
and the  Davidson algorithm\citep{DAVIDSON197587} is often a preferred choice for converging them. 
{Alternative algorithms include quasi Newton DIIS,\citep{hattig2000} Lanczos,\citep{Coriani2012} as well as shifted Davidson and GPLHR.\citep{Zuev2015,Vecharynski2016}}
The Davidson algorithm is nowadays routinely used to 
determine
both low-lying valence states and core and ionized 
states within the core-valence separation\citep{Cederbaum1980} (CVS) approximation.\citep{Coriani2015,Coriani2016erratum,Vidal2019} 
In the Davidson algorithm, residual vectors are added to a subspace in which the excited state equations are solved. As for the ground state, convergence is greatly improved by preconditioning the residual vector with the zeroth order Jacobian matrix. However, it has long been recognized that using a more accurate Jacobian as a preconditioner does \emph{not} necessarily improve the Davidson update; 
in the limit of an exact Jacobian, it gives no new search direction. Hence, the preconditioner must not be too good of an approximation of the Jacobian.\citep{OLSEN1990463, Helgaker2014} 
The Olsen algorithm removes this defect in the Davidson algorithm by requiring that the new search direction be orthogonal to the original eigenvector estimate.\citep{OLSEN1990463} 
For an exact approximation of the Jacobian matrix, the Olsen algorithm is equivalent to the Jacobi-Davidson\citep{sleijpen1995jacobi,sleijpen2000jacobi} algorithm. Both algorithms have been used in full  and multireference  configuration interaction theory.\citep{OLSEN1990463,VanDam1996,Leininger2001} 
In this paper, we use the Olsen estimate, or correction vector---derived in terms of a Newton update scheme---and apply the multimodel approximation to the equation for the estimate.
The Davidson, Olsen, and Jacobi-Davidson algorithms can all be viewed as Newton methods that incorporate ``subspace acceleration'' since the update estimates are used to build a reduced space in which the eigenvalue equation is solved.\citep{sleijpen1995jacobi,fokkema1998accelerated}

The 
update
estimates can either be added to a reduced space in which the excited state equations are solved, or they can be used in combination with DIIS. The latter is useful for the perturbative CC$n$ models ($n = 2, 3, \ldots$).\citep{hattig2000}  For these models, the equations of excitation order $n$ have analytical solutions, and, when those 
are inserted in the lower-order equations,  the excited state eigenvalue problem becomes nonlinear; 
the Jacobian transformation of an eigenvector becomes dependent on the corresponding eigenvalue.\citep{Christiansen1995,Koch1997} 
While useful, the DIIS method is not as robust for excited states as it is for the ground state. 
For 
inaccurate
starting guesses, 
it
may converge slowly 
and/or 
give
duplicate eigenvectors. 
As an alternative to DIIS, or as a preconvergence step,  one can use a nonlinear reduced space algorithm in which the space is repeatedly built until self-consistency.\citep{hattig2000}
One can also use 
reduced space algorithms 
directly on the 
non-folded CC$n$ equations,
but 
this implies increased storage requirements and hence
a stricter limit on the treatable system size.

Here we show that multimodel Newton algorithms can be applied successfully to the CC3 ground and excited state equations. 
Combining
CC3 ($\target$) 
with CCSD ($\aux$), 
we find that one may obtain both faster convergence and computational savings relative to algorithms based on the zeroth order Jacobian approximation. 
The algorithm we use for the ground state is a quasi Newton algorithm, accelerated by DIIS, where the CCSD Jacobian matrix is used to approximate the CC3 Jacobian. For excited states, we use a nonlinear reduced space Olsen algorithm where the 
CCSD Jacobian approximation is used
in 
the equations for the Olsen update estimates. 

\section{Multimodel Newton} \label{sec:multimodel_Newton}
\subsection{Ground state}
The coupled cluster ground state is parametrized as
\begin{align}
    \ket{\cc} = e^T \ket{\hf}, \quad  T = \sum_{\mu} t_{\mu} \tau_\mu, \label{eq:cc_state}
\end{align}
where $t_{\mu}$ and $\tau_{\mu}$ denote, respectively, the cluster amplitudes and the excitation operators with respect to the Hartree-Fock determinant $\ket{\hf}$. The state is determined by solving the amplitude equations
\begin{align}
    E_0 &= \Tbraket{\hf}{\bar{H}}{\hf} \\
    \Omega_{\mu} &= \Tbraket{\mu}{\bar{H}}{\hf} = 0, \label{eq:omega_standard}
\end{align}
where
\begin{align}
   \bar{H} = e^{-T} H e^T.
\end{align}
In Eqs.~\eqref{eq:cc_state}--\eqref{eq:omega_standard}, $\mu$ is restricted to a subset of excitations (singles, singles and doubles, \ldots), defining the standard hierarchy of coupled cluster models (CCS, CCSD, \ldots). The CC2 and CC3 methods are derived by partitioning the Hamiltonian as $H = F + V$, where $F$ is the Fock matrix, and performing a perturbation expansion of the highest-order amplitude equations in CCSD and CCSDT, respectively.\citep{Christiansen1995,Koch1997}

The exact Newton method may be derived as follows. First we expand the residual vector $\b{\Omega}$ about a guess $\b{t}_0$ for the amplitudes:
\begin{align}
    \Omega_\mu(\b{t}) = \Omega_\mu(\b{t}_0) + \sum_\nu \frac{\partial \Omega_\mu}{\partial t_\nu}\Big\vert_0 \Delta t_\nu + \order{\Delta \b{t}^2}. \label{eq:first_order_omega_1}
\end{align}
Here $\Delta \b{t} = \b{t} - \b{t}_0$. This expansion is conveniently expressed in matrix notation. By defining the coupled cluster Jacobian\citep{Koch1990}
\begin{align}
    A_{\mu\nu} = \frac{\partial \Omega_\mu}{\partial t_\nu} = \Tbraket{\mu}{[\bar{H}, \tau_\nu]}{\hf}, \label{eq:def_A}
\end{align}
we can rewrite Eq.~\eqref{eq:first_order_omega_1} as
\begin{align}
    \b{\Omega}(\b{t}) = \b{\Omega}(\b{t}_0) + \J(\b{t}_0) \Delta \b{t} + \order{\Delta \b{t}^2}. \label{eq:first_order_omega_2}
\end{align}
We wish to determine $\Delta \b{t}$ such that $\b{\Omega}(\b{t}) = 0$.
By truncating the expansion to first order and setting the result to zero, we obtain the Newton equation for $\Delta \b{t}$:
\begin{align}
    \b{0} = \b{\Omega}(\b{t}_0) + \J(\b{t}_0) \Delta \b{t}. \label{eq:Newton_equation}
\end{align}
The algorithm involves a set of macroiterations $n$ where
\begin{align}
    \b{t}^{n} = \b{t}^{n-1} + \Delta \b{t}^{n}, \quad n = 1,2,\ldots,
\end{align}
and
\begin{align}
    \b{0} = \b{\Omega}(\b{t}^{n-1}) + \J(\b{t}^{n-1}) \Delta \b{t}^{n} = \b{\rho}(\Delta \b{t}^{n}). \label{eq:Newton_equation_iteration}
\end{align}
To get the $n$th update estimate $\Delta \b{t}^{n}$, a series of microiterations is performed to solve Eq.~\eqref{eq:Newton_equation_iteration} approximately. 

In most cases, the Newton method converges rapidly in terms of macroiterations. Sufficiently close to the solution, the rate of convergence is even quadratic. However, because macro and microiterations are equally expensive, the Newton method is of limited use in practice without further approximations in Eq.~\eqref{eq:Newton_equation_iteration}. 

The standard zeroth order approximation\citep{Helgaker2014} is obtained  
by neglecting the fluctuation potential $V$
in Eq.~\eqref{eq:def_A}:
\begin{align}
    A_{\mu\nu} \approx (\b{\epsilon})_{\mu\nu} = \Tbraket{\mu}{[F,\tau_\nu]}{\hf} = \delta_{\mu\nu} \epsilon_\mu, \label{eq:A_orbital_differences}
\end{align}
where, for instance,
\begin{align}
    \epsilon_{ai} &= \epsilon_a - \epsilon_i \\
    \epsilon_{aibj} &= \epsilon_a + \epsilon_b - \epsilon_i - \epsilon_j,
\end{align}
and $\epsilon_p$ is the orbital energy of the $p$th canonical Hartree-Fock orbital.
Since $\b{\epsilon}$ is diagonal, Eq.~\eqref{eq:Newton_equation_iteration} becomes exactly solvable:
\begin{align}
    \Delta \b{t}^n = -\b{\epsilon}^{-1} \b{\Omega}^{n-1} \iff \Delta t_\mu^n = - \epsilon_\mu^{-1} \Omega_\mu^{n-1}. \label{eq:epsilon_t_estimate}
\end{align}
When combined with DIIS, this quasi Newton method often leads to rapid convergence.

{Poor convergence and numerical instabilities are expected if some of the orbital differences nearly vanish  (e.g., in bond dissociation processes). There has been some effort to alleviate these instabilities 
in the literature.\citep{Piecuch1994} Note that such problems are not unique to the orbital differences approximation in Eq.~\eqref{eq:A_orbital_differences} and that the same issues arise for other quasi Newton methods when the approximate Jacobian matrix is nearly singular.}

In DIIS, the $n$th amplitude estimate $\braket{\b{t}}^n$ is 
defined as
a linear combination of the previous and current Newton estimates, $\b{t}^m$, where $m = 1, 2, \ldots, n$:
\begin{align}
    \braket{\b{t}}^n = \sum_{m = 1}^n \lambda_m \b{t}^m = \sum_{m = 1}^n \lambda_m (\b{t}^{m-1} + \Delta \b{t}^{m}). \label{eq:diis_T}
\end{align} 
The 
weights $\lambda_m$ are determined by minimizing, in the least-squared sense, the averaged error vector 
\begin{align}
    \braket{\b{\Omega}}^n = \sum_{m = 1}^{n} \lambda_m \b{\Omega}^m \label{eq:diis_Omega}
\end{align}
under
the requirement that
\begin{align}
\sum_{m = 1}^n \lambda_m = 1. \label{eq:diis_lambda_eq_1}
\end{align}
The number of previous estimates included in these sums is typically restricted in practice. 
In our experience, eight previous estimates is normally sufficient for the ground state. 

We propose a
multimodel quasi Newton algorithm where the auxilliary model ($\aux$) is used to approximate the target model ($\target$) Jacobian. In other words, we determine the $n$th update estimate for the amplitudes, $\Delta \b{t}^{n}$, by solving the Newton equation
\begin{align}
    \b{0} = \b{\Omega}_{\target}(\b{t}^{n-1}) + \J_{\aux}(\b{t}^{n-1}) \Delta \b{t}^{n}. \label{eq:Newton_equation_iteration_mixed}
\end{align}
Here $\J_{\aux}$ is evaluated using the current guess $\b{t}^{n-1}$ for the target amplitudes, not amplitudes obtained from an $\aux$ calculation.
Note that each microiteration involves one transformation by $\J_{\aux}$, which has a cost of $\order{M^{k_\aux}}$. The only term whose cost scales as $\order{M^{k_\target}}$ is the evaluation of $\b{\Omega}_{\target}$, which is performed once per macroiteration.

\begin{algorithm}[H]
\caption{Ground state multimodel Newton with DIIS}
\begin{algorithmic}[1]
    \State \textbf{Input:} $\tau, \alpha$
    \State $\b{t} \gets \b{t}_0$
    \State $\b{\Omega}_{\target} \gets \b{\Omega}_{\target}(\b{t}_0)$
    \State $E, \Delta E \gets E(\b{t}_0)$
    \While {$\norm{\b{\Omega}_{\target}} > \tau$ or $\Delta E > \tau$}    
        \State \emph{Solve the Newton equation}
        \State \Call{LinEqDavidson}{$\Delta \b{t}$, $-\b{\Omega}_{\target}$, $\J_{\aux}$, $\alpha \norm{\b{\Omega}_{\target}}$, $0$}
        \State \emph{Use current and previous solutions to get DIIS estimate}
        \State $\b{t} \gets \b{t} + \Delta \b{t}$
        \State Get $\braket{\b{t}}$ by minimizing $\braket{\b{\Omega}_{\target}}$  
        \State $\b{t} \gets \braket{\b{t}}$
        \State \emph{Construct residual and energy}
        \State $\b{\Omega}_{\target} \gets \b{\Omega}_{\target}(\b{t})$
        \State $E \gets E(\b{t})$ \emph{and compute energy change} $\Delta E$
    \EndWhile
\end{algorithmic}
\end{algorithm}

Implementation details are given in Algorithm 1. The update estimates are obtained by using a linear equation Davidson algorithm to solve Eq.~\eqref{eq:Newton_equation_iteration_mixed} to within a relative threshold: 
\begin{align}
\norm{\b{\rho}} < \alpha \norm{\b{\Omega}_{\target}}.
\end{align}
The linear equation Davidson procedure is described in Algorithm 2. Upon convergence of the Newton equation, we apply DIIS as described in Eqs.~\eqref{eq:diis_T}--\eqref{eq:diis_lambda_eq_1} to get a new guess for the amplitudes. The process is repeated until $\norm{\b{\Omega}_{\target}} < \tau$ and $\Delta E < \tau$.

\begin{algorithm}[H]
\caption{Linear equation Davidson algorithm with $\b{\epsilon}$ preconditioner}
\begin{algorithmic}[1]
\Procedure{LinEqDavidson}{$\b{x}$, $\b{b}$, $\b{A}$, $\tau$, $\omega$}
        \State $\b{c} \gets (\b{\epsilon} - \omega)^{-1} \b{b}$
        \State $\b{c} \gets \b{c}/\norm{\b{c}}$
        \State $\mathcal{K} = \{ \b{c} \}; \; \mathcal{L} = \{ \J \b{c} \}$
        \While {$\norm{\b{\rho}} > \tau$}
            \State \emph{Update subpace arrays and solve reduced problem}
            \State $b^\mathrm{red}_{k} = \b{c}_k^T \b{b}, \quad \b{c}_k \in \mathcal{K}$
            \State $A^\mathrm{red}_{kl} = \b{c}_k^T \b{\sigma}_l, \quad \b{c}_k \in \mathcal{K}, \quad  \b{\sigma}_l \in \mathcal{L}$
            \State \emph{Get} $\b{x}^\mathrm{red}$ \emph{from} $ (\J^\mathrm{red} - \omega \b{I}) \b{x}^\mathrm{red} = \b{b}^\mathrm{red}$ 
            \State \emph{Construct next guess and associated residual}
            \State $\b{x} \gets \sum_k x^\mathrm{red}_k \b{c}_k$
            \State $\b{\rho} \gets -\b{b} + (\J - \omega \b{I}) \b{x}$ 
            \State \emph{Precondition residual and add it to $\mathcal{K}$}
            \If {$\norm{\b{\rho}} > \tau$}
            \State \emph{Set} $\b{c} \gets  (\b{\epsilon} - \omega)^{-1}\b{\rho}$ \emph{and orthonormalize it against $\mathcal{K}$}
            \State $\mathcal{K} \gets \mathcal{K} \cup \{\b{c}\}$
            \State $\mathcal{L} \gets \mathcal{L} \cup \{\J \b{c}\}$
            \EndIf
        \EndWhile
\EndProcedure
\end{algorithmic}
\end{algorithm}

The multimodel algorithm is straightforward to use when $\target$ and $\aux$ are dimension consistent. If $\b{\Omega}_{\target}$ in Eq.~\eqref{eq:Newton_equation_iteration_mixed} is an $m$-dimensional vector, then $\J_{\aux}$ must be an $m \times m$ matrix.
Thus, if we want to converge the folded CC2 equations ($\target$), then we can apply CCS as the auxilliary model ($\aux$). Similarly, if our target is the CCSD equations ($\target$), then full space CC2 ($\aux$) is applicable. Finally, for folded CC3 equations ($\target$), we can apply full space CC2 as well as CCSD ($\aux$).

If the dimensionalities of $\aux$ and $\target$ do not match, block matrices can be combined. In the case of CCSDT ($\target$), one could for instance use the auxilliary Jacobian
\begin{align}
    \J_\aux = \begin{pmatrix} \J_\mathrm{CCSD} & 0 \\ 0 & \epsilon_{\mu_3} \delta_{\mu_3\nu_3} \end{pmatrix}.
\end{align}

\subsection{Excited states} \label{sec:newton_excited}
In coupled cluster theory there are two approaches to excited states: response theory\citep{Koch1990} and equation of motion theory.\citep{Stanton1993} Both lead to the same eigenvalue problem:\citep{Koch1990,Stanton1993} 
\begin{align}
    \J \b{R}_n = \omega_n \b{R}_n, \quad n = 1, 2, 3, \ldots \label{eq:eigenvalue_problem}
\end{align} 
The eigenvalues $\omega_n$ are the excitation energies, and the eigenvectors $\b{R}_n$ define, together with the ground state multipliers, the excited states $\ket{\Psi_n}$ in equation of motion theory.\citep{Stanton1993}

 In the following we assume that we are solving for $\b{R}_n$. Of course, the described iteration scheme can be used for the left states as well, the only difference being that $\J$ is replaced by $\J^T$ and $\b{R}$ by $\b{L}$:
\begin{align}
    \b{L}_n^T \J  = \omega_n \b{L}_n^T, \quad n = 1, 2, 3, \ldots \label{eq:eigenvalue_problem_left}
\end{align}

To derive a Newton equation for a specific excited state, we let the current eigenvector guess be denoted as $\b{R}_0$ and define the residual at a nearby $\b{R}$ as
\begin{align}
   \b{\rho}(\b{R}) = \J \b{R} - \omega(\b{R}) \b{R}, \label{eq:rho_for_JD}
\end{align}
where the energy function $\omega$ is given by
\begin{align}
   \omega(\b{R}) = \frac{\b{R}_0^T \J \b{R}}{\b{R}_0^T \b{R}}. \label{eq:eigenvalue_estimate}
\end{align}
As noted by Sleijpen and Van der Vorst,\citep{sleijpen1995jacobi} choosing the fixed vector $\b{R}_0$ as the left vector in $\omega$ 
leads to a Newton equation whose solution is identical to the Jacobi-Davidson estimate, provided orthogonality to the initial guess is imposed. Let us show this.
By performing an expansion of the residual $\b{\rho}$ in Eq.~\eqref{eq:rho_for_JD} about $\b{R}_0$, that is,
\begin{align}
    \rho_\mu(\b{R}) = \rho_\mu(\b{R}_0) + \sum_\nu J_{\mu\nu}(\b{R}_0)   \Delta R_\nu + \order{\Delta\b{R}^2},
\end{align}
where
\begin{align}
   J_{\mu\nu}(\b{R}_0) = \frac{\partial \rho_\mu}{\partial R_\nu}\Big\vert_0,
\end{align}
we find
that
\begin{align}
   \b{J}(\b{R}_0) = \mathscr{P}_\perp(\b{R}_0) [\J - \omega(\b{R}_0)\b{I}].
\end{align}
Here $\mathscr{P}_\perp(\b{X})$ denotes the projector onto the orthogonal complement of  $\b{X}$:
\begin{align}
   \mathscr{P}_\perp(\b{X}) = 1 - \b{X}\b{X}^T, \quad \norm{\b{X}} = 1.
\end{align}
By truncating the expansion of $\rho(\b{R})$ at first order and setting the result to zero, we obtain 
\begin{align}
   \b{0} = \b{\rho}(\b{R}_0) + \mathscr{P}_\perp(\b{R}_0) [\J - \omega(\b{R}_0)\b{I}] \Delta \b{R}. \label{eq:es_redundant}
\end{align}
This equation does not have a unique solution. If $\Delta \b{R}$ is a solution to Eq.~\eqref{eq:es_redundant}, then so is
\begin{align}
    \Delta \b{R}(\gamma) = (\gamma - 1)\b{R}_0 + \gamma \Delta \b{R}, \quad \gamma \in \mathbb{R}. \label{eq:norm_red}
\end{align}
The non-uniqueness can be removed by requiring orthogonality with the initial guess,
\begin{align}
   \Delta \b{R}^T \b{R}_0 = 0,  \label{eq:es_DR_R_ortho}
\end{align}
but other options are possible.\citep{sleijpen1995jacobi}
Combining Eq.~\eqref{eq:es_redundant} with the orthogonality condition in Eq.~\eqref{eq:es_DR_R_ortho} defines a Newton method where $\Delta \b{R}$ is equal to the Jacobi-Davidson update estimate:\citep{fokkema1998accelerated,sleijpen1995jacobi,sleijpen2000jacobi}
\begin{align}
    \b{0} &= \b{\rho}(\b{R}^{n-1}) + \mathscr{P}_\perp(\b{R}^{n-1}) [\J - \omega(\b{R}^{n-1})\b{I}] \Delta \b{R}^n \label{eq:es_redundant_iterative} \\
    0 &= [\Delta \b{R}^n]^T \b{R}^{n-1}, \quad n = 1,2,3,\ldots \label{eq:es_DR_R_ortho_iterative}
\end{align}

The non-uniqueness of $\Delta \b{R}$ is lost if an approximation of $\J$ is introduced in the equation. However,
the method can be
generalized
to approximations by observing that Eq.~\eqref{eq:es_redundant} can be cast in the form
\begin{align}
    [\J - \omega(\b{R}_0)\b{I}] \Delta \b{R} = -\b{\rho}(\b{R}_0) + \beta \b{R}_0 \label{eq:newton_es_rewrite}
\end{align}
where $\beta$ is the $\b{R}_0$-prefactor arising from $\mathscr{P}_\perp(\b{R}_0)$. This suggests that if we instead use an approximation of $\J$ in Eq.~\eqref{eq:newton_es_rewrite}, we can adjust $\beta$ such that Eq.~\eqref{eq:es_DR_R_ortho} remains valid. For a general approximation $\b{B}$ of the target $\J_{\target}$, this can be achieved by
setting 
\begin{align}
    [\b{B} - \omega(\b{R}_0)\b{I}] \Delta \b{R} = -\b{\rho}_{\target}(\b{R}_0) + \beta \b{R}_0, \label{eq:newton_approximate_A}
\end{align}
with $\beta$ defined as
\begin{align} 
    \beta = \frac{\b{R}_0^T [\b{B} - \omega(\b{R}_0)\b{I}]^{-1} \b{\rho}_{\target}(\b{R}_0)}{\b{R}_0^T [\b{B} - \omega(\b{R}_0)\b{I}]^{-1} \b{R}_0}.
\end{align}
Thus, we get iterative scheme\citep{sleijpen2000jacobi}
\begin{align}
    \Delta \b{R}^{n} = - \b{Y}^n + \beta_n \b{W}^n, \quad n = 1,2,3,\ldots \label{eq:JacobiDavidsonEstimate}
\end{align}
where
\begin{align}
    [\b{B} - \omega(\b{R}^{n-1})\b{I}] \b{Y}^n &= \b{\rho}_{\target}(\b{R}^{n-1}) \label{eq:JD1} \\
    [\b{B} - \omega(\b{R}^{n-1})\b{I}] \b{W}^n &= \b{R}^{n-1} \label{eq:JD2}
\end{align}
and 
\begin{align} 
    \beta_n = \frac{[\b{R}^{n-1}]^T \b{Y}^n}{[\b{R}^{n-1}]^T \b{W}^n}.
\end{align}
The procedure described by Eqs.~\eqref{eq:JacobiDavidsonEstimate}--\eqref{eq:JD2} gives the update estimates of the Olsen algorithm.\citep{OLSEN1990463,saad2011numerical} In the multimodel approach, we let $\b{B} = \J_\aux$.

For accurate approximations of $\J_\target$, the Olsen estimate is superior to that of Davidson.
Expressing the Davidson estimate as
\begin{align}
    \Delta \b{R}_\mathrm{dav} = -(\b{B} - \omega_0 \b{I})^{-1} (\J_\target - \omega_0 \b{I}) \b{R}_0, \label{eq:Davidson_update}
\end{align}
we see that it becomes parallel to 
$\b{R}_0$ as $\b{B} \rightarrow \J_\target$.\citep{Helgaker2014,OLSEN1990463} 

We have implemented a nonlinear reduced space Olsen algorithm; see Algorithm 3. 
This algorithm can be used for the perturbative CC$n$ models, where the Jacobian transformation $\J_{\target}(\omega)$ depends on the eigenvalue $\omega$; in particular, it is an algorithm to solve the nonlinear eigenvalue problem
\begin{align}
    \J_{\target}(\omega) \b{R} = \omega \b{R},
\end{align}
where $\J_{\target}(\omega)$ is defined by inserting the highest-order part of Eq.\eqref{eq:eigenvalue_problem} into the lower-order parts of the equation (see, \emph{e.g.}, Ref.~\citenum{hattig2000}).
In the outer iterations of Algorithm 3, the residuals and energies are first constructed. Then, in order to get the next guess for the states, we perform a set of inner iterations in which a subspace consisting of Olsen updates is constructed. In the inner iterations, the energies are kept fixed, and the convergence threshold is chosen to be relative to the norm of the current residuals (given by $\alpha$). Note also that microiterations must be  performed in each inner iteration to determine the Olsen estimates (line {23}). Hence, the algorithm has three layers of iterations: outer, inner, and micro. Outer and inner iterations involve $\J_\target$ transformations; microiterations involve $\J_\aux$ transformations. When there are $k$ nonconverged states in an inner iteration, we have to solve $2k$ Olsen equations. These $2k$ equations are all solved in the same reduced space---using Algorithm 2---to within a relative threshold (given by $\alpha'$).

To transform the trial vectors, we have to specify an excitation energy.
If the vector is one of the states $\b{R}_n$, it is transformed by $\J_\target(\omega_n)$, where $\omega_n$ is the current energy estimate for $\b{R}_n$ (see line {6}). The vector is also transformed by $\J_\target(\omega_n)$ if it is the $\mathcal{K}$-orthonormalized Olsen estimate $\Delta \b{R}_n$ associated with $\b{R}_n$  (line {25}). Note that the energy dependence of $\J_\target$ is implicitly ignored when solving the reduced space eigenvalue equation
\begin{align}
    \J^\mathrm{red} \b{R}^\mathrm{red} = \lambda \,  \b{S}^\mathrm{red} \b{R}^\mathrm{red}.
\end{align}
{When orthonormalizing new trial vectors against $\mathcal{K}$ (line 25), the non-orthogonality of the initial trial vectors must be taken into account.}

A few observations can be made. If the residuals in the outer iterations are zero, then the residuals in the first inner iteration also vanish, ensuring self-consistency between inner and outer iterations. Furthermore, if an inner iteration solution is dominated by the initial guess and Olsen estimates corresponding to that root, the outer and inner residuals should be similar; since the trial vectors that dominate the state (line {17}) have all been transformed with the same energy estimate (the same $\omega$, line {27}), the final inner iteration residual (line 17) will approximate the subsequent outer residual (line {5}).

\begin{algorithm}[H]
\caption{Nonlinear multimodel Olsen}
\begin{algorithmic}[1]
    \State \textbf{Input:} $\tau, \alpha, \alpha'$
    \State $\b{R} \gets \b{R}_0$
    \While {$\norm{\b{\rho}_{\target}} > \tau$ or $\Delta \omega > \tau$}  
        \State \emph{Construct residuals and update energies $\omega$}
        \State $\b{\rho}_{\target} \gets \J_{\target}(\omega) \b{R} - \omega \b{R}$
        \State $\omega \gets \b{R}^T \J_{\target}(\omega) \b{R}$ \emph{and compute the energy change $\Delta \omega$}
        \State \emph{Initialize Olsen trials and transforms and begin inner iterations}
        \State $\mathcal{K} = \{ \b{R} \} $
        \State $\mathcal{L} = \{ \J_{\target}(\omega) \b{R} \}$
        \State $\lambda, \Delta \lambda \gets \omega$
        \While {{$\norm{\b{\rho}} > \alpha \cdot \mathrm{max} \norm{\b{\rho}_{\target}}$}}
            \State \emph{Update subpace arrays and solve reduced problem}
            \State $S^\mathrm{red}_{kl} = \b{c}_k^T \b{c}_l, \quad \b{c}_k, \b{c}_l \in \mathcal{K}$
            \State $A^\mathrm{red}_{kl} = \b{c}_k^T \b{\sigma}_l, \quad \b{c}_k \in \mathcal{K}$,   $\b{\sigma}_l \in \mathcal{L}$
            \State \emph{Get $\b{R}^\mathrm{red}$ and $\lambda$ from $\J^\mathrm{red} \b{R}^\mathrm{red} = \lambda \; \b{S}^\mathrm{red} \b{R}^\mathrm{red}$} 
            \State \emph{Construct guesses and residuals}
            \State $\b{R} \gets \sum_k R^\mathrm{red}_k \b{c}_k$
            \State $\b{\rho} \gets \sum_k R^\mathrm{red}_k \b{\sigma}_k  - \lambda \b{R}$
            \State \emph{For nonconverged states, get Olsen updates and add to $\mathcal{K}$}
            \If {$\norm{\b{\rho}} > \alpha \cdot \mathrm{max} \norm{\b{\rho}_{\target}}$}
            \State $\b{\rho} \gets \b{\rho} / \norm{\b{R}}$ and $\b{R} \gets \b{R} / \norm{\b{R}}$
            \State $\bar{\b{R}} \gets \b{R} \cdot \norm{\b{\rho}} $
            \State 
            \Call{LinEqDavidson}{$[\b{Y}, \b{W}]$, $[\b{\rho}, \bar{\b{R}}]$, $\J_{\aux}$, $\alpha'  \cdot \mathrm{min}\norm{\b{\rho}}$, $[\lambda, \lambda]$}   
            \State $\beta = \b{R}^T\b{Y}/\b{R}^T\b{W}$
            \State \emph{Set} $\Delta \b{R} =  -\b{Y} + \beta \b{W}$ \emph{and orthonormalize it against $\mathcal{K}$}
            \State $\mathcal{K} = \mathcal{K} \cup \{ \Delta \b{R}\}$
            \State $\mathcal{L} = \mathcal{L} \cup \{ \J_\target(\omega)\Delta \b{R} \}$
            \EndIf
        \EndWhile 
    \EndWhile
\end{algorithmic}
\end{algorithm}

Two modifications of Algorithm 3 are worth noting. First, if $\J_\target$ is independent of $\omega$, no inner iteration loop is necessary; one can simply expand the Olsen subspace until convergence is reached. The obtained algorithm is similar to the standard Davidson algorithm,\citep{DAVIDSON197587} the only  difference being that Olsen estimates, instead of preconditioned residuals, form the reduced space. This Olsen algorithm can be used when $\target$ is a standard model like CCSD or CCSDT. Second, we get a nonlinear $\b{\epsilon}$-Davidson algorithm by replacing the Olsen estimate $\Delta \b{R}$ with the Davidson estimate $(\b{\epsilon} - \lambda)^{-1} \b{\rho}$ in Algorithm 3. The resulting Davidson algorithm is very similar to that suggested by Hättig and Weigend.\citep{hattig2000}

{The Olsen and Davidson algorithms often reduce the outer iteration residual norms of the roots uniformly. However, this is not the case, for example, when restarting from converged roots and requesting additional roots, or when a new root is discovered late in the iterative procedure. In an inner iteration, only $\Delta \b{R}$ from non-converged $\b{R}$ are added to $\mathcal{K}$. However, the already converged roots are updated (line 17) and transformed in the subsequent outer iteration (line 5). 
To avoid unnecessary transformations, one can modify the algorithm such that converged roots are not updated and transformed in the subsequent outer iteration.
}

\begin{figure*}[t]
    \centering
    \includegraphics[trim={1.5cm 0.5cm 2cm 1cm},clip, width=\linewidth]{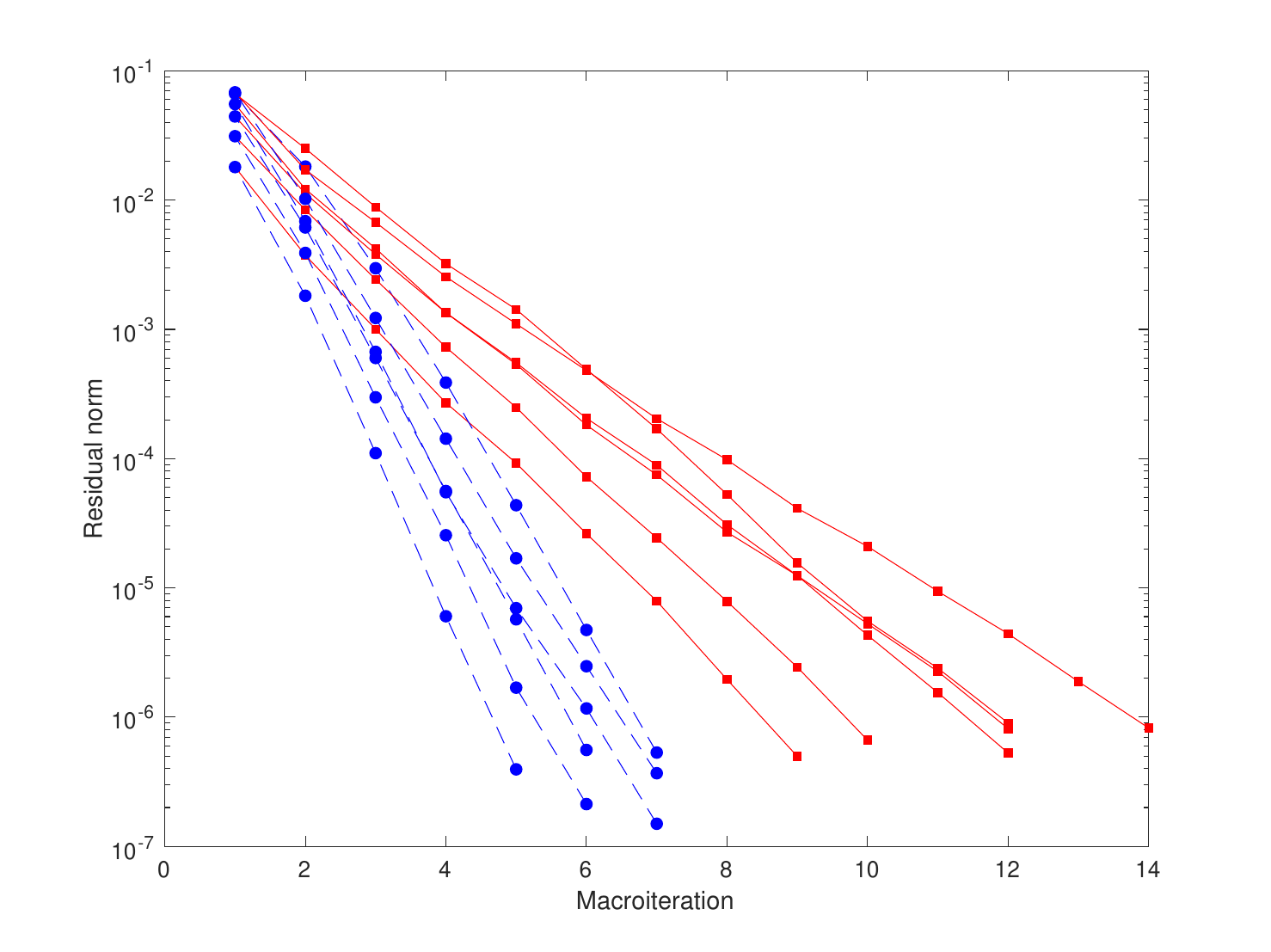}
    \caption{Convergence of ground states using $\b{\epsilon}$-Newton (red) and CC3/CCSD multimodel Newton (blue). Calculations were performed on water, formaldehyde, sulphur dioxide, acetamide, pyridazine, and cytosine with the aug-cc-pVDZ basis set.}
    \label{fig:ground_state}
\end{figure*}

\section{Results and discussion}

The multimodel Newton algorithms were implemented in a development version of the recently released electronic structure program eT.\citep{eT_paper} Here we restrict ourselves to the case of accelerating the convergence of CC3 equations ($\target$) using the CCSD Jacobian matrix ($\aux$) to approximate the CC3 Jacobian. 
For the ground state,
comparisons
are made 
between the standard  $\b{\epsilon}$-Newton approach, as  described by Eqs.~\eqref{eq:epsilon_t_estimate}--\eqref{eq:diis_lambda_eq_1}, and the multimodel Newton algorithm (Algorithm 1). 
In ground state calculations, eight vectors are used in the DIIS extrapolation. 
For excited states, we compare the nonlinear $\b{\epsilon}$-Davidson  and the nonlinear multimodel Olsen algorithms. The Olsen algorithm is described in Algorithm 3, and the $\b{\epsilon}$-Davidson algorithm is obtained from this algorithm by the modification described in Section \ref{sec:newton_excited}. In the reduced space algorithms, we use a maximum reduced space dimension of 100, resetting the space to the current solutions when the dimensionality is exceeded. {When converging six states simultaneously, the Olsen equations are solved with a maximum dimension equal to 120.} In all cases, the starting guesses are obtained from a CCSD calculation. 

The following thresholds are used. Unless otherwise specified, we use a residual threshold of $\tau = 10^{-6}$ to show the stability of the algorithms and their full convergence behavior.
The relative threshold is selected to be $\alpha = 10^{-2}$ for the ground state (see Algorithm 1). For the excited state, we use $\alpha = \alpha' = 10^{-1}$  (see Algorithm 3). 

\begin{table*}[t]
\caption{Total
wall time to converge the CC3 ground state. The total time is denoted as $T$ and the number of macroiterations by $N_\mathrm{it}$. $T_\mathrm{micro}$ is the percentage of time spent in the microiterations.
All calculations were performed on a node with two {Intel Xeon E5-2699 v4} processors using 44 threads and 1.5 TB shared memory.}
\begin{ruledtabular}
\centering
\begin{tabular}{llcccccc}
        & & & \multicolumn{2}{c}{$\b{\epsilon}$-Newton} 
        & \multicolumn{3}{c}{CC3/CCSD Newton} \\
        \cline{4-5} 
        \cline{6-8}
        & Basis & $M$ 
        & $T$
        & $N_\mathrm{it}$
        & $T$ 
        & $N_\mathrm{it}$
        & $T_\mathrm{micro}$ ($\%$) \\
    \hline 
    Water           & aug-cc-pVDZ & 41  & 1 s    & 9  & 1 s    & 5 & 40 \\
    Formaldehyde    & aug-cc-pVDZ & 64  & 4 s    & 10 & 4 s    & 6 & 26  \\
    Sulphur dioxide & aug-cc-pVDZ & 73  & 27 s   & 12 & 21 s   & 7 & 16 \\
                    & aug-cc-pVTZ & 142 & 5.0 m  & 13 & 3.3 m  & 7 & 17  \\
    Acetamide       & aug-cc-pVDZ & 137 & 4.1 m  & 12 & 2.4 m  & 6 & 16  \\
    Pyridazine      & aug-cc-pVDZ & 174 & 18.3 m & 12 & 12.2 m & 7 & 13 \\
    Cytosine        & aug-cc-pVDZ & 229 & 2.8 h  & 14 & 1.5 h  & 7 & 9  \\
\end{tabular}
\end{ruledtabular}
\label{tab:ground_state}
\end{table*}

For ground and valence excited states, we present convergence plots and wall times for water (H$_2$O), formaldehyde  (COH$_2$), sulphur dioxide (SO$_2$), acetamide (C$_2$H$_5$NO), pyridazine (C$_4$H$_4$N$_2$), and cytosine (C$_4$H$_5$N$_3$O). For core excited states, we present results for the oxygen edge of water (H$_2$O), formaldehyde (COH$_2$), and acetamide (C$_2$H$_5$NO) as well as the nitrogen edge of pyridazine (C$_4$H$_4$N$_2$), all calculated within the CVS approximation.\citep{Cederbaum1980,Coriani2015,Coriani2016erratum} The geometries are available from Ref.\citenum{geometries}.

We consider the ground state first;
see Figure \ref{fig:ground_state} for ground state convergence plots using 
$\b{\epsilon}$-Newton (red) and CC3/CCSD Newton (blue). The number of macroiterations needed in CC3/CCSD Newton is approximately cut in half compared to $\b{\epsilon}$-Newton. For quantitative results ($\norm{\b{\Omega}} < 10^{-4}$),  CC3/CCSD Newton requires 3--5 iterations while 5--9 iterations are required for $\b{\epsilon}$-Newton. This reduction implies significant computational savings. 

\begin{figure*}[t]
    \centering
    \includegraphics[trim={1.6cm 1cm 2cm 1cm},clip, width=\linewidth]{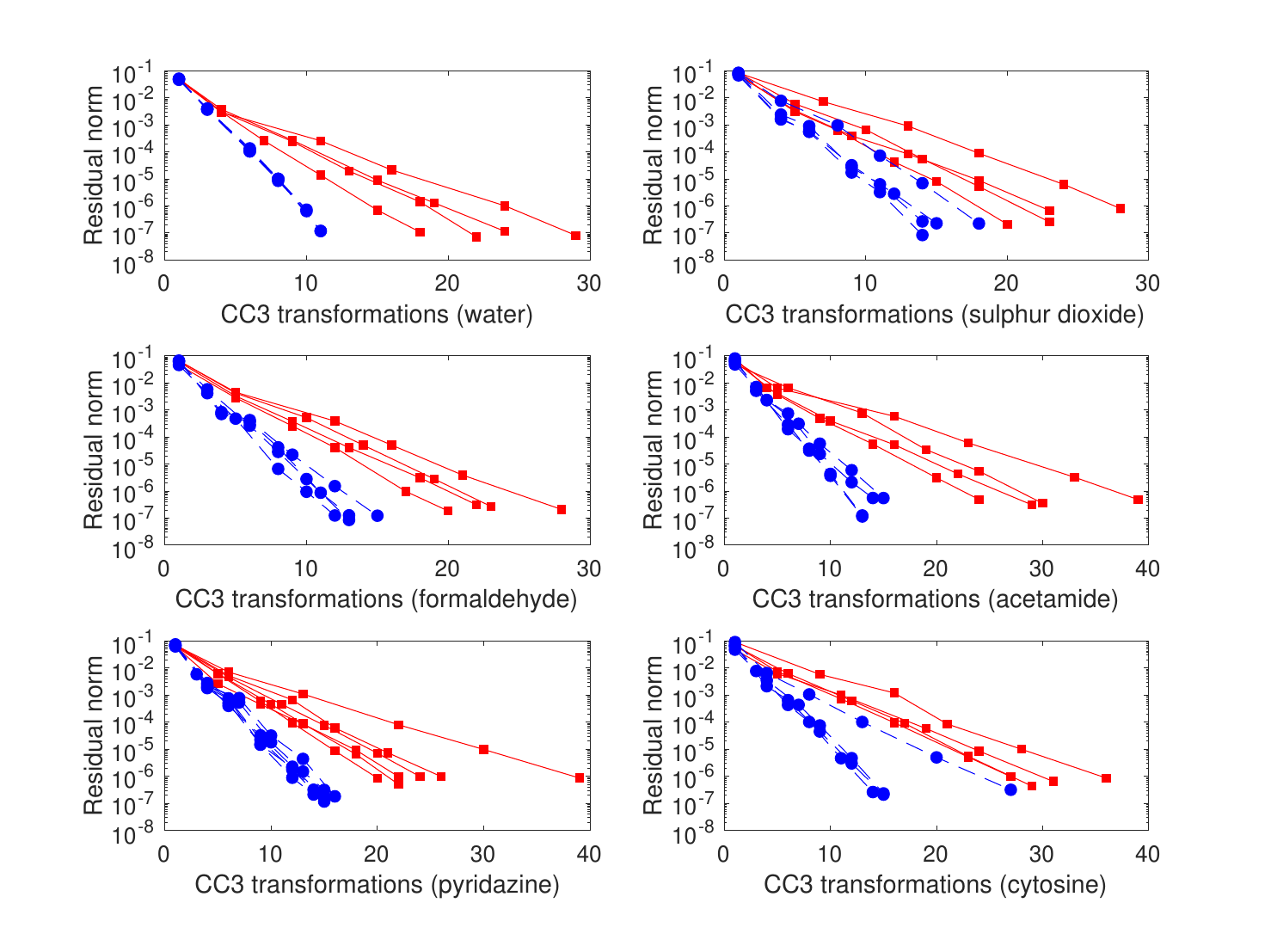}
    \caption{Convergence of valence excited states using the nonlinear $\b{\epsilon}$-Davidson (red) and nonlinear CC3/CCSD Olsen (blue) algorithms. The aug-cc-pVDZ basis set was used in all calculations.}
    \label{fig:valence_excited}
\end{figure*}

\begin{table*}[t]
\caption{Convergence of valence excited states in CC3. The total time to converge the excited states is denoted by $T$, the total number of CC3 transformations by $N_\mathrm{CC3}$, the number of  roots by $n_\mathrm{roots}$, and  the percentage of time spent doing CCSD microiterations by $T_\mathrm{micro}$. All calculations were performed on a node with two {Intel Xeon E5-2699 v4} processors using 44 threads and 1.5 TB shared memory.}
\begin{ruledtabular}
    \centering
    \begin{tabular}{llccccccc}
        & & & & \multicolumn{2}{c}{$\b{\epsilon}$-Davidson} 
        & \multicolumn{3}{c}{CC3/CCSD Olsen} \\         
        \cline{5-6} 
        \cline{7-9}
        & Basis 
        & $M$ 
        & $n_\mathrm{roots}$
        & $T$
        & $N_\mathrm{CC3}$
        & $T$ 
        & $N_\mathrm{CC3}$        
        & $T_\mathrm{micro}$ ($\%$) \\
    \hline
    Water           & aug-cc-pVDZ & 41  & 4 & 18 s   & 93  & 15 s   & 42 & 46 \\ 
    Formaldehyde    & aug-cc-pVDZ & 64  & 4 & 53 s   & 93  & 52 s   & 53 & 42 \\
                    & aug-cc-pVTZ & 138 & 4 & 10.1 m & 98  & 10.2 m & 57 & 43 \\
    Sulphur dioxide & aug-cc-pVDZ & 73  & 4 & 6.6 m  & 94  & 5.1 m  & 61 & 14 \\
                    & aug-cc-pVTZ & 142 & 4 & 57 m   & 97  & 43 m   & 62 & 15 \\
    Acetamide       & aug-cc-pVDZ & 137 & 4 & 62 m   & 122 & 36 m   & 55 & 22 \\
    Pyridazine      & aug-cc-pVDZ & 174 & 6 & 7.3 h  & 153 & 4.8 h  & 89 & 13 \\
    Cytosine        & aug-cc-pVDZ & 229 & 4 & 41 h   & 123 & 27 h   & 71 & 12 \\
    \end{tabular}
    \label{tab:excited_state}
\end{ruledtabular}
\end{table*}

In Table \ref{tab:ground_state} we list wall times for the ground state calculations. 
As expected, we find that the computational savings of CC3/CCSD Newton increase with the size of the system; 
the proportion of time spent performing CCSD microiterations should decrease as $M$ increases. In the case of cytosine/aug-cc-pVDZ ($M = 229$), the CCSD microiterations make up only 9\% of the total time, and since the amplitude equations converge in 7 macroiterations with CC3/CCSD Newton---compared to the 14 iterations needed with $\b{\epsilon}$-Newton---the wall time is nearly halved: from {$2.8$} to {$1.5$} hours. Savings are not only obtained for the largest systems; in Table \ref{tab:ground_state}, non-negligible savings are obtained for systems with more than about 100 orbitals.

In the case of valence excited states, we 
obtain
similar
improvements
in
the
convergence;
see Figure \ref{fig:valence_excited} for the convergence plots comparing nonlinear $\b{\epsilon}$-Davidson (red) and CC3/CCSD Olsen (blue).
Each subplot is for a given molecular system and shows the convergence of the four lowest-lying excited states as a function of the number of CC3 transformations. For a given state, we count the CC3 transformations required to form the residual and energy (lines {5 and 6} in Algorithm 3) and the transformations of Olsen trial vectors associated with the state (line {27}). 
The results are similar to those
obtained for the ground state:
the number of CC3 transformations is reduced by about  a factor of 
 two,
although the
variation 
is higher than for the ground state. 
Four roots were converged in all cases except pyradizine; for this system, we needed to converge two additional states to ensure that the algorithms converged the same set of states.

Wall times for the valence excited state calculations are given in Table \ref{tab:excited_state}. 
As for the ground state, 
computational savings 
are found to 
increase with the size of the system. For  cytosine/aug-cc-pVDZ ($M = 229$), the total wall time is reduced from {41} to {27} hours, and  the time spent doing CCSD microiterations make up only {12\%} of the total time. Although the savings generally increase with system size, we find that the CC3/CCSD Olsen algorithm provides non-negligible savings for systems with more than about 100 orbitals. 

 \begin{figure*}[t]
    \centering
    \includegraphics[trim={1.6cm 0.5cm 2cm 1cm},clip, width=\linewidth]{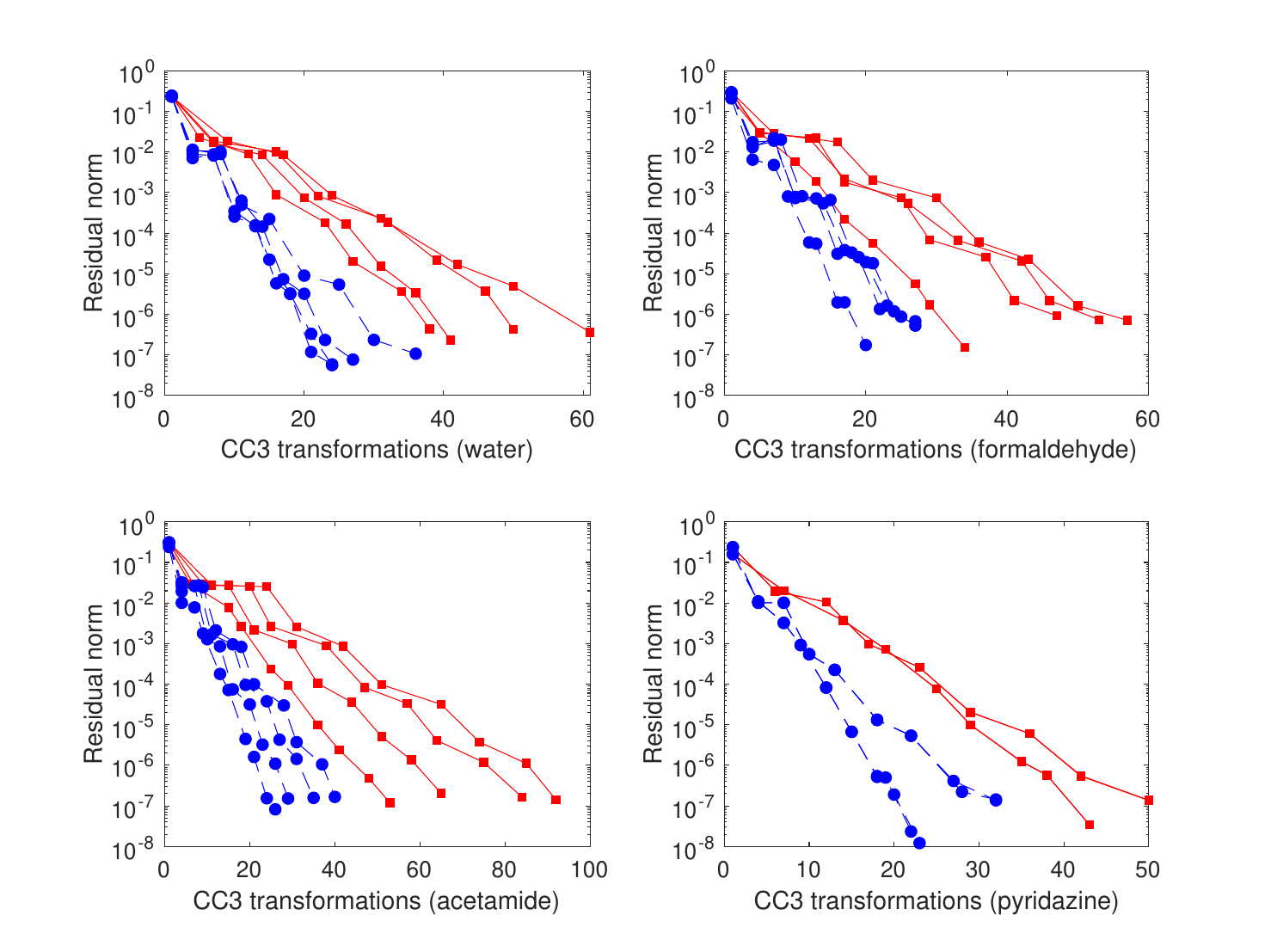}
    \caption{Convergence of CVS core excited states using the nonlinear $\b{\epsilon}$-Davidson (red) and nonlinear CC3/CCSD Olsen (blue) algorithms. The  aug-cc-pCVTZ basis was used on the excited cores and  aug-cc-pVDZ basis was used on the remaining atoms.}
    \label{fig:core_excited}
\end{figure*}

\begin{table*}[t]
\caption{Convergence of CVS core excited states in CC3. The total time to converge the excited states is denoted by $T$, the total number of CC3 transformations by $N_\mathrm{CC3}$, and  the percentage of time spent doing CCSD microiterations by $T_\mathrm{micro}$. On the excited cores (O, N), we use the aug-cc-pCVTZ basis. The aug-cc-pVDZ basis is used on the other atoms. All calculations were performed on a node with two {Intel Xeon E5-2699 v4} processors using 44 threads and 1.5 TB shared memory.}
\begin{ruledtabular}
    \centering
    \begin{tabular}{lcccccc}
        & & \multicolumn{2}{c}{$\b{\epsilon}$-Davidson} 
        & \multicolumn{3}{c}{CC3/CCSD Olsen} \\    
        \cline{3-4} 
        \cline{5-7}
         & $M$ 
         & $T$
         & $N_\mathrm{CC3}$
         & $T$ 
         & $N_\mathrm{CC3}$
         & $T_\mathrm{micro}$ ($\%$) \\
    \hline
    Water (O)        & 77  & 42 s   & 190 & 77 s   & 111 & 68 \\
    Formaldehyde (O) & 100 & 2.9 m  & 191 & 4.1 m  & 99  & 62 \\
    Acetamide (O)    & 173 & 1.3 h  & 294 & 1.3 h  & 130 & 52 \\
    Pyridazine (N)   & 246 & 10.9 h & 186 & 9.5 h  & 109 & 31
    \end{tabular}
    \label{tab:core_excited}
\end{ruledtabular}
\end{table*}

Convergence plots for core excited states are shown in Figure \ref{fig:core_excited}. Like for the ground and valence excited states, roughly half the number of macroiterations are needed in CC3/CCSD Olsen compared to $\b{\epsilon}$-Davidson. However, as is evident from the timings given  in Table \ref{tab:core_excited}, this reduction in macroiterations does not translate to significant computational savings. 
The implementation of CC3 core excitations has an iterative cost 
that scales as
$\order{M^6}$.\citep{Paul2019} 
For CCSD, the CVS approximation is implemented in eT through the projection technique described by Coriani and Koch:\citep{Coriani2015,Coriani2016erratum} the transformation by the CCSD Jacobian is first constructed, and then elements of the vector are set to zero if they do not correspond to an excitation out of the core orbital(s). Hence,
with the implementation used here, core excitations are not obtained at reduced scaling for CCSD.
The linear transformation by the CCSD CVS-Jacobian can 
of course
be implemented and used directly, as has been done by Vidal \emph{et al}.\citep{Vidal2019} This transformation has a theoretical scaling of $\order{M^5}$ (see the Supporting Information), implying an $M^6$/$M^5$ scaling for the $\target$/$\aux$ pair. In other words,
using the CCSD CVS-Jacobian transformation would likely yield significant savings also for core excited states. 

{In order to obtain some indication of the performance in more challenging situations, we consider the symmetric OH stretching of water (see Ref.~\citenum{Olsen1996} for the geometry) and determine the ground state and six valence excited states. The results are given in Table \ref{tab:h2o_stretch}, and show similar improvements in convergence. One exception is for excited states when the OH length is $R = 2.0 R_e$. Relatively small changes are observed in the time spent in microiterations. We should also mention that roundoff errors in CC3/CCSD Olsen are becoming noticeable as $R$ increases; for $R = 2.5 R_e$, we find a run-to-run variation---in 10 calculations using four threads---of a few transformations relative to the 189 obtained in the first calculation (185--190). This variation can be reduced by solving the Olsen equations more accurately, that is, by lowering $\alpha'$ in Algorithm 3.}

\begin{table*}[t]
\caption{Convergence of ground and six valence excited CC3 states at different OH bond lengths ($R$) in \ce{H2O}/cc-pVDZ. The equilibrium bond length is denoted $R_e$. The number of CC3 transformations is denoted by $N_\mathrm{CC3}^x$, where $x = \mathrm{gs}, \mathrm{es}$ refers to the ground and excited state. The time spent performing microiterations is similarly denoted $T_\mathrm{micro}
^x$. All calculations were performed on an Intel Core i7-8086K processor using one thread.}
\begin{ruledtabular}
    \centering
    \begin{tabular}{lcccccc}
        & \multicolumn{2}{c}{$\b{\epsilon}$-Newton/Davidson} 
        & \multicolumn{4}{c}{CC3/CCSD Newton/Olsen} \\ 
                \cline{2-3} 
        \cline{4-7}
        $R$ & $N_\mathrm{CC3}^\mathrm{gs}$ & $N_\mathrm{CC3}^\mathrm{es}$ &  $N_\mathrm{CC3}^\mathrm{gs}$ &
        $T_\mathrm{micro}^\mathrm{gs}(\%)$ &
        $N_\mathrm{CC3}^\mathrm{es}$ &
        $T_\mathrm{micro}^\mathrm{es}(\%)$\\
    \hline
    $1.0 \, R_e$ & 8  & 106 & 5 & 38 & 61  & 52  \\
    $1.5 \, R_e$ & 11 & 130 & 5 & 44 & 80  & 47  \\
    $2.0 \, R_e$ & 14 & 132 & 7 & 57 & 122 & 44  \\
    $2.5 \, R_e$ & 14 & 242 & 8 & 50 & 189 & 45
    \end{tabular}
\end{ruledtabular} \label{tab:h2o_stretch}
\end{table*}

\section{Concluding remarks}
In this work, we have shown that the convergence of the CC3 ground and excited state equations can be improved by approximating the Jacobian in Newton-type methods at the CCSD level. 
By using a DIIS-accelerated quasi Newton algorithm for the ground state, and a nonlinear Olsen algorithm for excited states, convergence was achieved in about half the number of CC3 iterations---as compared to algorithms based on the standard zeroth order approximation of the Jacobian.
The improved convergence 
of these multimodel CC3/CCSD 
algorithms 
is 
found to give computational savings for both ground and valence excited states---also for 
molecular
systems that are considered small in the context of CC3. Similar improvements in convergence are found for 
core excited states
within the CVS approximation. 
However, this did not lead to savings in computational time,
since the implementations 
of
the CCSD Jacobian and CC3 CVS-Jacobian transformations both  scale as $\order{M^6}$.
This can 
most
likely
be remedied by 
using an
implementation of the $\mathcal{O}(M^5)$ CCSD CVS-Jacobian transformation\citep{Vidal2019} instead of the $\mathcal{O}(M^6)$ projection technique\citep{Coriani2015,Coriani2016erratum} employed in this work.

\section{Supplementary material}
Includes a derivation of the $\mathcal{O}(M^5)$ scaling of the CCSD-CVS Jacobian transformation.

\section{Data availability statement}
The geometries used are available at {10.5281/zenodo.3753153} (Ref.~\citenum{geometries}).
Other files are available from the corresponding author upon request.

\begin{acknowledgments}
We thank Rolf H.~Myhre {and Alexander C. Paul} for useful discussions. We acknowledge computing resources through UNINETT Sigma2, the National Infrastructure for High Performance Computing and Data Storage in Norway (project number NN2962k) and funding from the Marie Sk{\l}odowska-Curie European Training Network ``COSINE - COmputational Spectroscopy In Natural sciences and Engineering'', Grant Agreement No. 765739, and the Research Council of Norway through FRINATEK projects 263110 and 275506.
\end{acknowledgments}

\bibliography{main}

\end{document}